\documentstyle[12pt,fleqn]{article}
\vspace*{2cm}
\def\jmp{J.~Math.~Phys.}
\def\PL{Phys. Lett.}

\def\PTP{Prog. Theor. Phys.}
\def\MP{Int. Journ. Mod. Phys. A}

 \def\nn{\nonumber}
 
 \def\br{\begin{flushright}}
 \def\er{\end{flushright}}
 \def\be{\begin{equation}}
 \def\ee{\end{equation}}
 \def\bea{\begin{eqnarray}}
 \def\eea{\end{eqnarray}}
 \def\ba{\begin{array}}
 \def\ea{\end{array}}
 \def\ben{\begin{eqnarray}}
 \def\een{\end{eqnarray}}
 \def\pd{{\mbox {$\partial$}}}
 \def\cD{{\mbox {${\cal D}$}}}
 \def\cF{{\mbox {${\cal F}$}}}
 \def\cL{{\mbox {${\cal L}$}}}
 \def\cA{{\mbox {${\cal A}$}}}
 \def\cH{{\mbox {${\cal H}$}}}
 \def\ga{{\mbox {${\alpha}$}}}
 \def\ep{{\mbox {${\epsilon}$}}}
 
 \def\dab{{\mbox {${\delta_{\alpha\beta}}$}}}
 
 \def\tphi{{\mbox {${\tilde \phi}$}}}
 
 \def\tH{{\mbox {${\tilde H}$}}}
 
 \def\YMH{Yang-Mills-Higgs$\;$}

\begin{document}
\thispagestyle{empty}
\baselineskip=18pt
\br
DPNU-95-27
\er
\vspace{5mm}
\begin{center}
  {\large\bf NON-COMMUTATIVE DIFFERENTIAL GEOMETRY}\par
\vspace{2ex}
  {\large\bf AND STANDARD MODEL}\\
\vspace{20mm}
{\Large\sf Katsusada Morita and Yoshitaka Okumura$^{*)}$}\\
\vspace{3ex}
{\sf
  Department of Physics,
  Nagoya University, Nagoya, 464-01}\\
\vspace{2ex}
{\sf  $^{*)}$ Department of Natural Sciences,
  Chubu University, Kasugai, Aichi, 487}\\
\vspace{2ex}
\vspace{8ex}
 \begin{large}
{\sf Abstract}\\
 \end{large}
\end{center}
\vspace{10mm}
We incorporate Sogami's idea in the standard model into our previous
formulation of non-commutative differential geometry by extending
the action of the extra exterior derivative operator on spinors
defined over the discrete space-time $M_4\times Z_2$.
The extension consists in making
it possible to require that the operator
become nilpotent when acting on the spinors.
It is shown that the generalized field strength
leads to the most general, gauge-invariant
\YMH lagrangian
even if
the extra exterior derivative operator
is not nilpotent, while the fermionic part remains intact.
The proof is given for a single Higgs model.
The method is applied to reformulate the standard model by
putting left-handed fermion doublets on the upper sheet
and right-handed fermion singlets on the lower sheet
with generation mixing among quarks being taken into account.
We also present a matrix calculus of the method without referring
to the discrete space-time.
\vfill
\newpage
\baselineskip=18pt
\parskip=6pt
\section*{\large \bf {{\mbox \S 1} Introduction}}
%
The standard model of elementary particles has passed all experimental
checks so far and there is no doubt as to its validity
up to energies explored by present accelerators.
Nonetheless, it is annoyed with
many fundamentally unknown parameters
and, moreover, it essentially relies on the little-understood
Higgs mechanism.
It is a common belief in particle physicists
that the standard model is a low energy effective
theory of more fundamental theory.
But, what is the more fundamental theory?
There are many attempts to break the present deadlock in particle physics
by invoking interesting physical motivations and/or mathematical
apparatus.\par
Among them there is one making use of quite unfamiliar
mathematics. It is Connes' gauge theory$^{1)}$ which
aims at geometrizing
\YMH broken gauge theory
in terms of his non-commutative geometry (abbreviated NCG hereafter).
The fact that the successful particle model is a broken gauge theory
led Connes to envisage fine structure of the space-time,
which allows to naturally introduce symmetry breaking into the gauge theory
by tensoring Dirac operators on continuous and finite spaces.
The simplest such spaces are $M_4$, our 4-dimensional
minkowski space-time,
and $Z_2$, two-points space.
Connes' theory has since been expounded by various authors$^{2)-9)}$.
\par
In this paper we introduce Connes' approach
to the readers using new version of non-commutative differential
geometry based on the underlying space-time
$X=M_4\times Z_2$.
The old version was proposed in Ref.9)
by modifying Sitarz' formalism$^{8)}$
and will be up-graded below by incorporating Sogami's clever idea$^{10)}$
in the standard model.
The most prominent feature of
the new version is to
start from fermions
in accord with the assumption$^{1)}$ that
the underlying fields in NCG
are the spinor fields.
In this respect our formalism and
Sitarz' one further developed by Ding et al.$^{8)}$
are similar but alike only in appearance.
We shall point out the differences more concretely in the text.
We hope the present version will help the readers to understand
Connes' gauge theory more easily dispensing with abstract
Connes' mathematics.
It is
worthwhile studying NCG in simpler setting.
\par
The plan of this paper goes as follows.
In the next section we introduce the new formulation of
differential calculus on $X$.
It is an up-grade version of that formulated in Ref.9).
We shall consider in \S3
gauge theory over $X$ and prove that
it is nothing but \YMH broken gauge theory
with a single Higgs field.
The method will be applied to the standard model in \S4.
We essentially reproduce the results of Ref.12).
We shall present a matrix version of the method
in the chiral space in
\S5 without referring to the discrete space-time.
The last section is devoted to discussions.
Two Appendices are included containing some remarks on our algebraic
manipulations.
\section*{\large \bf {{\mbox \S 2}
New Differential Calculus on Discrete
Space-Time $X$}}
\par
Let $\psi(x,y)$ be a spinor defined on the discrete space-time
$X=M_4\times Z_2$, where $x \in M_4$ and
$y=\pm$ denote two elements of $Z_2$.
(We regard $Z_2$ as two-points space but not as
a discrete group.)
With respect to the internal symmetry
it is assumed to gauge-transform as
\be
\psi(x,y)\rightarrow ^g\!\!\psi(x,y)=\rho(g(x,y))\psi(x,y),
\label{eqn:a1}
\ee
where $g(x,y)$ is a local gauge function belonging to
the gauge group $G_y$ and $\rho$ indicates the fermion representation (rep)
of the gauge group. Note that $G_+\not=G_-$, in general.
By convention we assume that left-handed fermions are placed on the upper
sheet labeled by $y=+$, whilst right-handed fermions are put on
the lower sheet labeled by $y=-$.
Consequently, we write $\psi(x,+)=\psi_L(x)$ and $\psi(x,-)=\psi_R(x)$,
where $\psi_L(x)=P_+\psi(x)$ and $\psi_R(x)=P_-\psi(x)$
for a Dirac spinor $\psi(x)$
with chiral projection operators
$P_{\pm}={1\over 2}(1\pm\gamma_5)$
satisfying $P_+^2=P_+, P_-^2=P_-$ and $P_+P_-=P_-P_+=0$.
\par
We next introduce the generalized
exterior derivative operator ${\bf d}=d+d_\chi$
acting on $\psi(x,y)$, where
$d$ is the ordinary exterior derivative operator and
$d_\chi$ turns out to describe symmetry breaking
in the theory.
In order to present detailed construction of the broken gauge theory
in the next section
it is convenient to make basic definitions in this section.
The operator ${\bf d}$ is defined by
\be\ba{cl}
&{\bf d}\psi(x,y)=d\psi(x,y)+d_\chi\psi(x,y),\\[2mm]
&d\psi(x,y)=\pd_\mu\psi(x,y) d{\hat x}^\mu,\\[2mm]
&d_\chi\psi(x,y)=[M(y)\psi(x,-y)+iC(y)\psi(x,y)]\chi,\;\;
M(-y)=M^{\dag}(y),\\[2mm]
&d^2{\hat x}^\mu=d_\chi d{\hat x}^\mu=d\chi=d_\chi\chi=0,
\;\;{\hat x}^\mu:\;{\rm dimensionless}\;{\rm coordinates}.
\label{eqn:a2}
\ea\ee
Here
the basis of the ``cotangent space''
of $X$ is denoted by $\{d{\hat x}^\mu, \chi\}$, which is
dual to the basis $\{\partial_\mu, \partial_\chi\}$
of the ``tangent space'' of $X$
with, for instance,
$\chi(\partial_\chi)=1$ (in mass dimension one)\footnote{The
``derivation'' $\partial_\chi$ kinematically
``generates'' the fermion mass.
To see this let us consider free Dirac lagrangian in the following form.
$\cL_D^0(x)=i{\bar\psi}(x)\gamma^\mu\partial_\mu\psi(x)
-{\bar\psi}(x)M\psi(x)=
i{\bar\psi}_L(x)\gamma^\mu\partial_\mu\psi_L(x)
+i{\bar\psi}_R(x)\gamma^\mu\partial_\mu\psi_R(x)
-{\bar\psi}_L(x)M\psi_R(x)-{\bar\psi}_R(x)M\psi_L(x).$
Defining
$\partial_\chi\psi_L(x)=M\psi_R(x)+ic_L\psi_L(x)$,
$\partial_\chi\psi_R(x)=M\psi_L(x)+ic_R\psi_R(x)$
and putting $\psi(x,+)=\psi_L(x)$ and $\psi(x,-)=\psi_R(x)$,
we get
$\cL_D^0(x)=\sum_{y=\pm}[i{\bar\psi}(x,y)\gamma^\mu\partial_\mu\psi(x,y)
-{\bar\psi}(x,y)\partial_\chi\psi(x,y)]$,
where we have made use of the relations
${\bar\psi}_L\psi_L={\bar\psi}_R\psi_R=0$.
The operator $d_\chi$ acts on the spinor $\psi(x,y)$
as $d_\chi\psi(x,y)=(\partial_\chi\psi(x,y))\chi$,
which takes the form of the third equation of
Eq.(\ref{eqn:a2}) provided $M(+)=M(-)=M, C(+)=c_L$ and $C(-)=c_R$.
It will be shown in the next section that
gauge fields arise from the covariantization of $\partial_\mu$
and Higgs fields from that of $\partial_\chi$.}.
We assume that
$dd_\chi+d_\chi d=0$, hence ${\bf d}^2=d_\chi^2$ because $d^2=0$.
The symbol $\chi$ was first introduced by Sitarz$^{8)}$
in relation to Connes' NCG$^{1)}$.
We continue to employ it
although we drastically differ from Sitarz' formalism
as emphasized in Ref.9). For instance, we assume $d_\chi\chi=0$
in contrast with Sitarz' assumption
$d_\chi\chi=2\chi\wedge\chi\not=0$ \footnote{The authors in
Ref.11) claim that the relation
$d_\chi\chi=2\chi\wedge\chi$
is one of important characteristics in NCG.
On the contrary, we shall see below that
the assumption $d_\chi\chi=0$ is quite consistent with NCG.
The point is that there exist more than one definitions of the
action of $d_\chi$ on the 0-form (and the spinor)
since it is no longer a differential but  difference operator.}.
The action $d_\chi\psi(x,y)$
contains two matrix-valued functions $M(y)$ and $C(y)$, both of which
are assumed to be $x$-independent.
The case $C(y)=0$ reduces to the previous
definition (I-31)\footnote{We refer to Eq.(31) in Ref.9)
as Eq.(I-31).}.
Hence $M(y)$
is identical with the previous
one$^{9)}$, i.e., we assume a linear map $M(y):\cH_{-y}\rightarrow
\cH_{y}$, where $\cH_{\pm y}$ denote Hilbert spaces of spinors $\psi(x,\pm y)$.
We also assume a linear map $C(y):\cH_y\rightarrow \cH_{y}$.
A possible presence of the term $C(y)$ in the definition for
$d_\chi\psi(x,y)$ is suggested from the 2$\times $2 matrix formulation$^{12)}$
of Sogami's method$^{10)}$.
The precise role thereof in our present formulation will be clarified
in the next section.
\par
To distinguish linear maps from $\cH_{y}$ to
$\cH_{\pm y}$
we introduce the concept of grading.
We assign even grade to $C(y)$
and odd grade to $M(y)$. In general, linear maps depend on $x$,
so that we have both even and odd functions $f(x,y)$.
By definition we should consider only the products
$f(x,y)\psi(x,y')$
with $y'=(-1)^{\partial f}y$, where $\partial f=0$
for even function $f$ and $\partial f=1$
for odd function $f$ \footnote{
The degree $\partial f$ of the grade is defined up to mod 2.}.
Such products should be consistently calculable by the
usual matrix multiplication rule.
Furthermore Leibniz rule for the derivatives must also be applicable.
As for the ordinary exterior derivative $d$
there is no problem:
\be
d(f(x,y)\psi(x,y'))=(df(x,y))\psi(x,y')+f(x,y)(d\psi(x,y')).
\label{eqn:Leib1}
\ee
On the other hand, the extra exterior derivative $d_\chi$
is assumed to satisfy the graded Leibniz rule
\be
d_\chi(f(x,y)\psi(x,y'))=
(d_\chi f(x,y))\psi(x,y')+(-1)^{\partial f}f(x,y)(d_\chi\psi(x,y')).
\label{eqn:a3}
\ee
The reason why we should have extra factor $(-1)^{\partial f}$
in the above equation even for 0-form $f(x,y)$
will become clear in the next section.
According to the definition (\ref{eqn:a1})
we easily find
\be\ba{cl}
&d_\chi(f(x,y)\psi(x,y'))=[M(y)f(x,-y)\psi(x,-y')+
iC(y)f(x,y)\psi(x,y')]\chi\\[2mm]
&\qquad\qquad\qquad\qquad\,=[d_\chi(f(x,y))\psi(x,y')\\[2mm]
&\qquad\qquad\qquad\qquad\,
+(-1)^{\partial f}f(x,y)\{M(y')\psi(x,-y')+iC(y')\psi(x,y')\}\chi].
\label{eqn:a4}
\ea\ee
Assuming that $\partial_\chi f$ has opposite grade to that of $f$ \footnote{For
any functions $f,g$
we have $\pd (fg)=\pd f+\pd g$ mod 2.},
Eq.(\ref{eqn:a4}) yields
\be
\chi\psi(x,y)=\psi(x,-y)\chi,
\label{eqn:a5}
\ee
\be
d_\chi f(x,y)=[M(y)f(x,-y)-(-1)^{\partial f}f(x,y)M(y')]\chi,
\label{eqn:a6}
\ee
and
\be
C(y)f(x,y)=(-1)^{\partial f}
f(x,y)C(y').
\label{eqn:a7}
\ee
\par
Equation (\ref{eqn:a6}) was previously$^{9)}$ proposed (see, Eq.(I-1)).
By assumption (\ref{eqn:a1})
the gauge function $\rho(g(x,y))$
is even
so that
it commutes with $C(y)$ from Eq.(\ref{eqn:a7}):
\be
C(y)\rho(g(x,y))=\rho(g(x,y))C(y).
\label{eqn:a8}
\ee
Taking $f(x,y)=M(y)$ in Eqs.(\ref{eqn:a6}) and
(\ref{eqn:a7}) and remembering that $\pd M(y)=1$
we have
\be\ba{cl}
d_\chi M(y)&=2M(y)M(-y)\chi\\[2mm]
C(y)M(y)&=-M(y)C(-y).
\label{eqn:a9}
\ea\ee
As we showed in Ref.9),
consistent calculability based on Eq.(\ref{eqn:a6})
implies that
even functions $f(x,+)$ and $f(x,-)$ are
square matrices of dimensions $m$ and $n$, respectively,
while
odd functions $f(x,+)$ and $f(x,-)$ are matrices of types
$(m, n)$ and $(n, m)$, respectively.
Consequently, $\psi(x,+)$ is $m$-component spinor
and $\psi(x,-)$ $n$-component spinor regarding the internal
symmetry groups $G_+$ and $G_-$, respectively.
This presents a strong correlation
between rep contents of fermions and bosons (gauge {\it and} Higgs fields).
We also require the graded Leibniz rule for the product of
linear maps $f(x,y)$ and $f'(x,y')$ with
$y'=(-1)^{\partial f}y$
\be
d_\chi(f(x,y)f'(x,y'))=
(d_\chi f(x,y))f'(x,y')+(-1)^{\partial f}f(x,y)(d_\chi f'(x,y')),
\label{eqn:a10}
\ee
which leads to Sitarz' relation like Eq.(\ref{eqn:a5})
\be
\chi f(x,y)=f(x,-y)\chi.
\label{eqn:a11}
\ee
Conversely, if we assume Eq.(\ref{eqn:a11}),
we can prove the graded Leibniz rule (\ref{eqn:a10}) from Eq.(\ref{eqn:a6}).
The proof was given in the Appendix A of Ref.9).
It is impossible to exaggerate that
the relations (\ref{eqn:a5}) and (\ref{eqn:a11}) as they stand
are not matrix equations.
In particular,
we consider$^{9)}$ an algebraic sum $A+B\chi$ of two matrices $A$ and $B$
of, in general, different types, as far as the variable $y$
is explicit,
while Sitarz$^{8)}$
assume $A$ and $B$ to be of the same type.
Ding et al.$^{8)}$ assumed Eqs.(\ref{eqn:a5}) and (\ref{eqn:a11})
as matrix equations
and ended up with the commutativity and anticommutativity
of $\chi$ with bosonic and fermionic variables, respectively.
In their formalism
with $Z_2$ being taken as a symmetry group
consisting of elements $\{e,r=(CPT)^2;r^2=e\}$,
therefore,
$\chi$ is a commuting or anticommuting
``scalar''
in the matrix multiplication law.
\par
If we
employ the 2$\times$2 matrix rep
of elements of the $Z_2$-graded algebra$^{14)}$,
we can combine Eqs.(\ref{eqn:a5}) and (\ref{eqn:a11})
into matrix equations by considering both $y=\pm$ cases
simultaneously.
More about this in the Appendix A.
\par
According to Eq.(\ref{eqn:a2}) the operator $d_\chi$ is not diagonal
in the sense that $d_\chi\psi(x,y)$ contains both even and odd
functions $C(y)$ and $M(y)$.
The operator $d_\chi^2$ becomes diagonal, however,
since the function $[M(y)M(-y)-C(y)C(y)]$ is even:
\be
d_\chi^2\psi(x,y)=[M(y)M(-y)-C(y)C(y)]\psi(x,y)\chi\wedge\chi.
\label{eqn:a12}
\ee
Similarly, we obtain$^{9)}$ from Eq.(\ref{eqn:a6})
that
\be
\!\!\!\!\!\!\!\!\!\!
d_\chi^2f(x,y)=[M(y)M(-y)f(x,y)-f(x,y)M(y')M(-y')]\chi\wedge\chi,\;
y'=(-1)^{\partial f}y.
\label{eqn:a12-1}
\ee
This suggests that it is possible to assume the nilpotency of the
operator $d_\chi$
by putting
$M(y)M(-y)=C(y)C(y)$ \footnote{It is interesting to emphasize that
it is possible to realize the nilpotency of $d_\chi$
without Sitarz'
assumption $d_\chi \chi=2\chi\wedge\chi$.
This observation suggests itself that
realization of Connes' NCG in terms of the new symbol $\chi$
is not unique.
In other words, there exist various definitions of $d_\chi$ though
$d_\chi$ is uniquely determined on the whole algebra
once its action
on the 0-form (or the spinor) is defined.
See the second footnote on p.4.}.
This requirement precisely corresponds to
the condition $M^2=C^2$ as observed in Ref.13)
where it was shown that
Sogami's
method$^{10)}$
is equivalent to one version$^{14)}$ of NCG.
In the present formulation we shall not need
the nilpotency of $d_\chi$ and
make use of Eq.(\ref{eqn:a12})
in the next section even if $M(y)M(-y)\not=C(y)C(y)$.
However, it is interesting to remark that
the present formalism is applicable to both
nilpotent and non-nilpotent cases.
In this respect, too, we are differing from
Ref.8) where the nilpotency of $d_\chi$
is the central requirement.
\section*{\large \bf {{\mbox \S 3}
Gauge Theory on Discrete
Space-Time $X$}}
\par
Although the present formalism is applicable to
both global and local symmetries,
we consider exclusively local gauge theories in this article.
It is apparent that
${\bf d}\psi(x,y)$ is not covariant under local gauge transformations
(\ref{eqn:a1}).
It is covariantized by the familiar recipe:
\be
\cD(x,y)\psi(x,y)=({\bf d}+(\rho_{*}{\bf A})(x,y))\psi(x,y),
\label{eqn:a13}
\ee
where
$\rho_{*}$ is the differential rep for the fermions
and
the generalized gauge potential
\be
{\bf A}(x,y)=A(x,y)+\Phi(x,y)\chi,\;\;A(x,y)=A_\mu(x,y)d{\hat x}^\mu,
\label{eqn:a14}
\ee
is subject to the inhomogeneous gauge transformation
\be
^g\!{\bf A}(x,y)=g(x,y){\bf A}(x,y)g^{-1}(x,y)
+g(x,y){\bf d}g^{-1}(x,y).
\label{eqn:a15}
\ee
The notation $\rho_{*}$ means
$(\rho_{*}{\bf A})(x,y)=(\rho_{*}A)(x,y)+\Phi(x,y)\chi$, which
transforms like Eq.(\ref{eqn:a15})
with $g(x,y)\rightarrow \rho(g(x,y))$.
In particular, we have
\be
^g\Phi(x,y)=g(x,y)\Phi(x,y)g^{-1}(x,-y)
+g(x,y)\partial_\chi g^{-1}(x,-y).
\label{eqn:a15-1}
\ee
This inhomogeneous transformation law,
valid also for $\rho(g(x,y))$
in place of $g(x,y)$,
will be rederived in the Appendix B in a different way
and is entirely different from that proposed in Ref.8).
It stems from the different definitions of the operator $\partial_\chi$.
Our $\Phi(x,y)$ is a genuine {\it shifted} Higgs field, but
the scalar field in the generalized one-form in Ref.8)
represents unshifted Higgs field, although the scalar field itself transforms
inhomogeneously unless gauge transformations
defined at $e$ and $r$ in the notation of Ding et al.$^{8)}$
are identical to each other.
It follows that, if we define the {\it back-shifted}$^{\,\,3)}$ Higgs field
\be
H(x,y)=\Phi(x,y)+M(y),
\label{eqn:a18}
\ee
Eq.(\ref{eqn:a15-1}) implies the homogeneous
transformation law for it:
\be
^g\!H(x,y)=g(x,y)H(x,y)g^{-1}(x,-y).
\label{eqn:a19}
\ee
Hence, $M(y)$ determines VEV of the Higgs field
$<H(x,y)>=M(y)$.
The gauge group $G_+\times G_-$
is broken down to $H_+\times H_-$, where
$d_\chi h^{-1}(x,y)=0$ for
$h(x,y)\in H_y$.
The
matrix $M(y)$
determines$^{9)}$ the scale and the pattern of spontaneous symmetry breaking.
Higgs field enters into the theory
under the different guise in Sitarz' formalism.
\par
By construction $\cD(x,y)\psi(x,y)$ is gauge-covariant under Eq.(\ref{eqn:a1})
and can be rewritten as
\be\ba{cl}
\cD(x,y)\psi(x,y)&=[D(x,y)+d_\chi+\Phi(x,y)\chi]\psi(x,y)\\[2mm]
&=D(x,y)\psi(x,y)+[H(x,y)\psi(x,-y)+iC(y)\psi(x,y)]\chi,
\label{eqn:a16}
\ea\ee
where
\be
D(x,y)=d+(\rho_{*}A)(x,y).
\label{eqn:a17}
\ee
\par
The presence of the extra one-form basis $\chi$
prevents us from representing the covariant derivative (\ref{eqn:a13})
or (\ref{eqn:a16})
in terms of Clifford algebra.
Instead we introduce the associated spinor one-form
\be
{\tilde \psi}(x,y)=\gamma_\mu\psi(x,y)d{\hat x}^\mu
-i\alpha^{-2}\psi(x,y)\chi
\label{eqn:a20}
\ee
for some constant $\alpha^{-2}$.
Dirac $\gamma$-matrices are taken to satisfy
$\gamma_\mu\gamma_\nu+\gamma_\nu\gamma_\mu=2g_{\mu\nu}$
with ${\gamma_\mu}^{\dag}=\gamma_0\gamma_\mu\gamma_0$ and
$\gamma_5=-i\gamma_0\gamma_1\gamma_2\gamma_3.$
The Dirac lagrangian is then computed by the inner product
\be\ba{cl}
\cL_D(x,y)&=i<{\tilde\psi}(x,y),\cD(x,y)\psi(x,y)>\\[2mm]
&={\bar\psi}(x,y)i\gamma^\mu D_\mu(x,y)\psi(x,y)
-{\bar\psi}(x,y)H(x,y)\psi(x,-y)\\[2mm]
&\quad-i{\bar\psi}(x,y)C(y)\psi(x,y),
\label{eqn:a21}
\ea\ee
where we have defined the inner products of spinor one-forms through
\be\ba{cl}
&<\psi(x,y)d{\hat x}^\mu,\psi'(x,y)d{\hat x}^\nu>=
{\bar\psi}(x,y)\psi'(x,y)g^{\mu\nu}\\[2mm]
&<\psi(x,y)\chi,\psi'(x,y)\chi>=
{\bar\psi}(x,y)\psi'(x,y)<\chi,\chi>
\label{eqn:a22}
\ea\ee
with vanishing other inner products, and we have put
\be
<\chi,\chi>=-\alpha^2.
\label{eqn:a23}
\ee
Here the overbar of the spinor dictates the Pauli adjoint:
${\bar\psi}(x,y)=\psi^{\dag}(x,y)\gamma^0$.
By our assumption of the assignment of spinors
${\bar\psi}(x,y)C(y)\psi(x,y)=0$ for $y=\pm$ because of the relations
$P_+P_-=P_-P_+=0$. Thus $C(y)$
disappears from the Dirac lagrangian (\ref{eqn:a21})
and is nothing but Sogami's term$^{10)}$,
called $c_{L,R}$ in Ref.12).
The total Dirac lagrangian is the sum over $y=\pm$:
\be\ba{cl}
\cL_D(x)&=\sum_{y=\pm}\cL_D(x,y)\\[2mm]
&=\sum_{y=\pm}[{\bar\psi}(x,y)i\gamma^\mu D_\mu(x,y)\psi(x,y)
-{\bar\psi}(x,y)H(x,y)\psi(x,-y)],
\label{eqn:a24}
\ea\ee
which is hermitian since $H(x,-y)=H^{\dag}(x,y)$.
\par
Up to now we have concentrated on the fermionic sector.
The bosonic sector is characterized by
the generalized field strength
\be\ba{cl}
\cF(x,y)&=[{\bf d}+(\rho_{*}{\bf A})(x,y)]
\wedge[{\bf d}+(\rho_{*}{\bf A})(x,y)]\\[2mm]
&=\cD(x,y)\wedge\cD(x,y)\\[2mm]
&=[D(x,y)+d_\chi+\Phi(x,y)\chi]\wedge
[D(x,y)+d_\chi+\Phi(x,y)\chi].
\label{eqn:a25}
\ea\ee
Since ${\bf d}$
is not necessarily nilpotent,
it differs from
\be
{\bf F}(x,y)={\bf d}(\rho_{*}{\bf A})(x,y)+
(\rho_{*}{\bf A})(x,y)\wedge(\rho_{*}{\bf A})(x,y),
\label{eqn:a25-1}
\ee
which is not gauge covariant unless ${\bf d}^2=0$.
To evaluate Eq.(\ref{eqn:a25})
we need the expressions for
$d_\chi (\rho_{*}A)_\mu(x,y)$ and $d_\chi\Phi(x,y)$
\footnote{In addition to the usual differential calculus
we assume $d\wedge\chi+\chi\wedge d=0,
d_\chi\wedge d{\hat x}^\mu+d{\hat x}^\mu\wedge d_\chi=0,
d_\chi\wedge \chi-\chi \wedge d_\chi=0$.
The last relation implies $d_\chi\wedge (\chi\psi)=
(d_\chi\chi)\psi+\chi\wedge d_\chi\psi=\chi\wedge d_\chi\psi$
because $(d_\chi\chi)=0$ by assumption.}.
We determine them by requiring the graded Leibniz rule
for the generalized one-forms (\ref{eqn:a14}):
\be
{\bf d}({\bf A}(x,y)\wedge{\bf B}(x,y))=
({\bf d}{\bf A}(x,y))\wedge{\bf B}(x,y)
-{\bf A}(x,y)\wedge({\bf d}{\bf B}(x,y)).
\label{eqn:a26}
\ee
This is necessary to show the gauge covariance of Eq.(\ref{eqn:a25}).
We proved in the Appendix  A of Ref.9) that
Eq.(\ref{eqn:a26}) is valid provided that the ordinary
gauge field $A_\mu(x,y)$ is even, while
the shifted Higgs field $\Phi(x,y)$ is odd, which is consistent
with the assumption $\partial M(y)=1$ in view of Eq.(\ref{eqn:a18}).
The factor $(-1)^{\partial f}$ in Eq.(\ref{eqn:a10})
is essential to this proof.
This therefore fixes the operation
$d_\chi$ on $(\rho_{*}A)_\mu(x,y)$
and $\Phi(x,y)$ according to Eq.(\ref{eqn:a6}).
Substituting $d_\chi^2=[M(y)M(-y)-C(y)C(y)]\chi\wedge\chi$
from Eq.(\ref{eqn:a12})
into Eq.(\ref{eqn:a25})
we finally find that
\be
\cF(x,y)=F(x,y)+DH(x,y)\wedge\chi+(H(x,y)H(x,-y)-C(y)C(y))\chi\wedge\chi,
\label{eqn:a27}
\ee
where we put
\be
F(x,y)=D(x,y)\wedge D(x,y)={1\over 2}F_{\mu\nu}(x,y)d{\hat x}^\mu
\wedge{\hat x}^\nu,
\label{eqn:a33}
\ee
and assume $d{\hat x}^\mu\wedge\chi=-\chi\wedge d{\hat x}^\mu$
to obtain
\be
DH(x,y)=D(x,y)H(x,y)-H(x,y)D(x,-y)=(D_\mu H(x,y))d{\hat x}^\mu.
\label{eqn:a28}
\ee
The bosonic lagrangian
is then given by the sum\footnote{The fact that
the covariant derivative $\cD(x,y)$ contains both gauge and Higgs
fields which couple to fermions,
gives sever restriction on the form of the bosonic lagrangian (\ref{eqn:a30}),
although the number of parameters in the bare lagrangian
should be the same as that for renormalizable theory
unless some hidden symmetry is present.}
\be\ba{cl}
\cL_B(x)&=\sum_{y=\pm}\cL_B(x,y),\\[2mm]
\cL_B(x,y)&=-{\rm tr}{1\over 4g_y^2}<\cF(x,y),\cF(x,y)>,
\label{eqn:a30}
\ea\ee
where ${1\over 4g_y^2}$ is a coupling-constants-matrix
commuting with the gauge
transformation $\rho(g(x,y))$
and tr indicates the trace over the internal symmetry matrices.
Here the inner products are to be evaluated through
\be\ba{cl}
&<d{\hat x}^\mu\wedge d{\hat x}^\nu,d{\hat x}^\rho\wedge d{\hat x}^\sigma>
=g^{\mu\rho}g^{\nu\sigma}-g^{\mu\sigma}g^{\nu\rho}\\[2mm]
&<d{\hat x}^\mu\wedge \chi,d{\hat x}^\nu\wedge \chi>
=g^{\mu\nu}(-\alpha^2)\\[2mm]
&<\chi\wedge \chi,\chi\wedge \chi>={\beta^2\over 2},
\label{eqn:a31}
\ea\ee
while other inner products of basis two-forms are vanishing.
\par
To summarize we find the formula
\be\ba{cl}
\lefteqn{\cL_B(x)=\sum_{y=\pm}[-{1\over 2}{\rm tr}{1\over 4g_y^2}
F_{\mu\nu}(x,y)F^{\mu\nu}(x,y)
+\alpha^2{\rm tr}{1\over 4g_y^2}
(D_\mu H(x,y))^{\dag}D^\mu H(x,y)}\\[2mm]
&\qquad\quad\;\,\,-{1\over 2}\beta^2{\rm tr}{1\over 4g_y^2}
(H(x,y)H(x,-y)-C(y)C(y))^2].
\label{eqn:a32}
\ea\ee
Since $C(y)$ is gauge invariant because of Eq.(\ref{eqn:a8}),
the bosonic lagrangian (\ref{eqn:a32})
is the most general, gauge invariant \YMH lagrangian
provided there exists only one Higgs field.
The total lagrangian, the sum of
Eqs.(\ref{eqn:a24}) and (\ref{eqn:a32}),
becomes identical with
that obtained in the second reference of
Ref.8) provided $C(+)=C(-)$ and $\beta^2=2\alpha^4$.
Consequently,
it is not necessary to regard $Z_2$ as a discrete
group composed of $\{1,(CPT)^2\}$.
This group structure enforces the identification
$\psi(x,y)=-\psi(x,-y)$ which contradicts with
our assignment $\psi(x,+)=\psi_L(x)$ and
$\psi(x,-)=\psi_R(x)$.
The latter is more close to Connes' assignment$^{1)}$.
\par
The result (\ref{eqn:a32})
is markedly different from our previous ones (I-19)
and (I-11) which contain not-necessarily gauge-invariant
term $M(y)M(-y)$ in the Higgs
potential. The latter disappears from the scene by the introduction
of the term $C(y)$ in the definition for $d_\chi\psi(x,y)$.
Thus in the present formalism
we are not forced to
discard gauge-noninvariant
term tr$(H(x,+)H(x,-)-M(+)M(-))^2$ of Model I in Ref.9) by hand.
It is simply replaced with gauge-invariant one
tr$(H(x,+)H(x,-)-C(+)C(+))^2$.
Nevertheless, the minimum of the Higgs potential should occur
at $H(x,y)=M(y)$.
\section*{\large \bf {{\mbox \S 4}
Application to Standard Model}}
\par
In this section we apply the previous formalism
to reformulate the standard model taking the generation mixing among quarks
into account.
\par
To this end let us first recall that the Dirac lagrangian
for the standard model is given by the sum of leptonic and quark parts
\be
\cL_D=\cL_D^{(l)}+\cL_D^{(q)}.
\label{eqn:a34}
\ee
To fix the notation
we recapitulate it.
Writing the
weak lepton doublets and singlets, respectively,
in the $i$-th generation, as
$l_L^i=\left(
     \ba{cl}
     \nu_{eL}^i\\
     e_L^i\\
     \ea
     \right)$ and
$\;e_R^i$, the leptonic part reads
\be\ba{cl}
\cL_D^{(l)}=&\sum_i{\bar l_L^i}i\gamma^\mu
(\pd_\mu-i{g\over 2}{\vec\tau}\cdot
{\vec A}_\mu+i{g'\over 2}B_\mu)l_L^i \\[2mm]
&+\sum_i{\bar e_R^i}i\gamma^\mu(\pd_\mu+ig'B_\mu)e_R^i
-\sum_{i,j}[a_{ij}^{(e)}{\bar l_L^i}\phi e_R^j+
{a_{ji}^{(e)}}^{*}{\bar e_R^i}\phi^{\dag} l_L^j],
\label{eqn:a34-1}
\ea\ee
where
${\vec A}_\mu$ and $B_\mu$ are $SU(2)$ and $U(1)$
gauge fields, respectively, with corresponding gauge coupling constants
$g$ and $g'$, $\tau_i (i=1,2,3)$ Pauli matrices,
$\phi$ stands for Higgs doublet
      $\left(
      \ba{cl}
      \phi^+ \\
      \phi^0 \\
      \ea
      \right)$ and
$a_{ij}^{(e)}$ represent the Yukawa coupling constants.
It is possible to diagonalize the matrix $a^{(e)}=(a_{ij}^{(e)})$
in the generation space without changing gauge interactions
of leptons
as far as neutrinos are assumed to be massless.
On the other hand, the quark sector is defined by the lagrangian
\be\ba{cl}
\cL_D^{(q)}=&\sum_{\alpha,\beta,i}{\bar q}_L^{\alpha i}
i\gamma^\mu(\pd_\mu\dab
-i{g_s\over 2}(\lambda^a)_{\alpha\beta}G^a_\mu
-i{g\over 2}{\vec\tau}\cdot
{\vec A}_\mu\dab-i{g'\over 6}B_\mu\dab)q_L^{\beta i} \\[2mm]
&+\sum_{\alpha,\beta,i}{\bar d}_R^{\alpha i}i\gamma^\mu(\pd_\mu\dab
-i{g_s\over 2}(\lambda^a)_{\alpha\beta}G^a_\mu+i{g'\over 3}B_\mu\dab)
d_R^{\beta i} \\[2mm]
&+\sum_{\alpha,\beta,i}{\bar u}_R^{\alpha i}i\gamma^\mu(\pd_\mu\dab
-i{g_s\over 2}(\lambda^a)_{\alpha\beta}G^a_\mu
-i{2g'\over 3}B_\mu\dab)u_R^{\beta i} \\[2mm]
&-\sum_{\alpha,i,j}[a_{ij}^{(d)}
({\bar q}_L^{\alpha i}\phi)d_R^{\alpha j}+
{a_{ji}^{(d)}}^{*}{\bar d}_R^{\alpha i}(\phi^{\dag}q_L^{\alpha j})] \\[2mm]
&-\sum_{\alpha,i,j}[a_{ij}^{(u)}
({\bar q}_L^{\alpha i}{\tilde\phi})u_R^{\alpha j}+
{a_{ji}^{(u)}}^{*}{\bar u}_R^{\alpha i}({\tilde\phi}^{\dag}q_L^{\alpha j})],
\label{eqn:a34-2}
\ea\ee
where $\tilde\phi=i\tau_2\phi$,
left-handed quark doublet and right-handed quark singlets
in $i$-th generation with color $\ga$
are designated by
$q_L^{\alpha i}=\left(
             \ba{cl}
             u_L^{\alpha i}\\
             d_L^{\alpha i}\\
             \ea
             \right)$,
and $u_R^{\alpha i}$ and $d_R^{\alpha i}$, respectively,
$G_\mu^a (a=1,2,\cdots,8)$ stands for the gluon fields,
$\lambda^a (a=1,2,\cdots,8)$ are the Gell-Mann matrices,
$g_s$ denotes the QCD coupling constant and
$a_{ij}^{(q)}, q=u,d,$ represent Yukawa coupling matrices.
We assume $N_g$ generations, $i,j=1,2,\cdots,N_g.$
It is possible to diagonalize the matrices $a^{(q)}=(a_{ij}^{(q)})$
by biunitary transformations, i.e.,
$U^{(q)}a^{(q)}V^{(q)\dag}=
g^{(q)}$ are chosen to be
diagonal matrices with real, positive eigenvalues for
some unitary matrices $U^{(q)}$ and $V^{(q)}$.
Then, the gauge interactions of quarks
are written in terms of mass eigenstates,
where Kobayashi-Maskawa matrix $U=U^{(u)}U^{(d)-1}$ appears
for the charged current interactions.
In what follows we prefer to use gauge eigenstates as
exhibited in Eq.(\ref{eqn:a34-2}).
The non-diagonal matrices $a^{(q)}, q=u,d$
indicate the generation mixing among quarks.
\par
Sogami$^{10)}$ proposed to derive the bosonic
lagrangian from the sum of Eqs.(\ref{eqn:a34-1}) and
(\ref{eqn:a34-2}) and obtained a
constrained standard model\footnote{Sogami's reconstruction of the standard
model lagrangian in the bosonic sector is
quite different from that of Connes' NCG although the constraints
are more or less similar. Our aim is to greatly simplify
Connes' NCG in relation to Sogami's method.}.
Subsequently, the constraints are removed
by noting$^{12)}$ that
Sogami's method$^{10)}$ allows more parameters than originally supposed.
Then, the bosonic lagrangian is also given by the sum
\be
\cL_B=\cL_B^{(l)}+\cL_B^{(q)}.
\label{eqn:a34-3}
\ee
The relative weight between the leptonic and quark
contributions in Eq.(\ref{eqn:a34-3})
is determined only phenomenologically in the tree level.
\par
In the present formalism where $\psi(x,+)=\psi_L(x)$
and $\psi(x,-)=\psi_R(x)$ we should place the left-handed leptons {\it and}
quarks on the upper sheet, while
the right-handed leptons {\it and}
quarks are to be put on the lower sheet.
This needs a reconsideration of the derivation of Eq.(\ref{eqn:a34-3}).
\par
It is not difficult to
cast the sum of Eqs.(\ref{eqn:a34-1}) and (\ref{eqn:a34-2})
into the form (\ref{eqn:a24})
by choosing
\be\ba{cl}
\psi^i(x,+)&=\left(
             \ba{cl}
             l_L^i(x)\\
             q_L^{\alpha i}(x)\\
             \ea
             \right)\\[2mm]
\psi^i(x,-)&=\left(
             \ba{cl}
             e_R^i(x)\\
             d_R^{\alpha i}(x)\\
             u_R^{\alpha i}(x)\\
             \ea
             \right).
\label{eqn:a35}
\ea\ee
In what follows we shall omit
the generation and color indices.
Since, in each generation, the left-handed fermions are flavor doublets,
the right-handed fermions singlets and
quarks exist in three colors, $\alpha=R,B,G$,
$\psi(x,+)$ is 8-component spinor regarding the gauge group
$G_+=U(2)\otimes SU(3)$,
and
$\psi(x,-)$ consists of 7 components for the gauge group
$U(1)\otimes SU(3)$.
Consequently, the ordinary covariant derivatives
$D(x,+)=d+\cA(x,+)$ and $D(x,-)=d+\cA(x,-)$
are 8$\times$8 and 7$\times$7
matrices, respectively,
while
$H(x,+)=\Phi(x,+)+M(+)$ is 8$\times$7 matrix
and $H(x,-)=H^{\dag}(x,+)$ 7$\times$8 matrix.
This is valid for every generation.
In addition we should consider the generation space matrix,
either unit matrix or Yukawa coupling matrices as direct products
which are to be understood in the following expressions.
\par
Denoting $p$-dimensional
unit matrix by
$1_p$
and looking at Eqs.(\ref{eqn:a34-1}) and (\ref{eqn:a34-2}),
the assignment (\ref{eqn:a35})
gives
\be\ba{cl}
\cA(x,+)&=-{ig\over 2}A(x)\otimes 1_4
-{ig'\over 2}B(x)Y_L+1_2\otimes(-{ig_s\over 2}G(x))\\[2mm]
\cA(x,-)&=
-{ig'\over 2}B(x)Y_R-{ig_s\over 2}{\tilde G}(x)
\label{eqn:a37}
\ea\ee
and
\be
H(x,+)=
\left(
\ba{ccc}
a^{(e)}\phi(x)&0&0\\
0&a^{(d)}\phi(x)\otimes 1_3&a^{(u)}\tphi(x)\otimes 1_3\\
\ea
\right),
\label{eqn:a38}
\ee
where
$A(x)=\tau_iA^i_\mu(x)d{\hat x}^\mu, B(x)=B_\mu(x)d{\hat x}^\mu$,
$G(x)=
\left(
\ba{cc}
0&0\\
0&\lambda_aG^a_\mu(x)d{\hat x}^\mu\\
\ea
\right)$ and\hfill\break
${\tilde G}(x)=
\left(
\ba{ccc}
0&0&0\\
0&\lambda_aG^a_\mu(x)d{\hat x}^\mu&0\\
0&0&\lambda_aG^a_\mu(x)d{\hat x}^\mu\\
\ea
\right)$.
The hypercharge matrices for fermions  are denoted by
\be\ba{cl}
Y_L&=\left(
    \ba{cc}
    -1\cdot 1_2&0\\
    0&{1\over 3}\cdot 1_2\otimes 1_3\\
    \ea
    \right)\\[2mm]
Y_R&=\left(
    \ba{ccc}
    -2&0&0\\
    0&-{2\over 3}\cdot 1_3&0\\
    0&0&{4\over 3}\cdot 1_3\\
    \ea
    \right).
\label{eqn:a39}
\ea\ee
\par
Our next task is to evaluate the generalized field strength
$\cF(x,y)$ of Eq.(\ref{eqn:a27}).
The ordinary field strength $F(x,y)$ takes the form
\be\ba{cl}
F(x,+)&=-{ig\over 2}f(x)\otimes 1_4
-{ig'\over 2}f^0(x)Y_L+1_2\otimes(-{ig_s\over 2}G^{(L)}(x))\\[2mm]
F(x,-)&=
-{ig'\over 2}f^0(x)Y_R-{ig_s\over 2}G^{(R)}(x),
\label{eqn:a40}
\ea\ee
where
\be\ba{cl}
f(x)&={1\over 2}{\vec\tau}\cdot
{\vec f}_{\mu\nu}d{\hat x}^\mu\wedge d{\hat x}^\nu,\\[2mm]
f^0(x)&={1\over 2}f^0_{\mu\nu}d{\hat x}^\mu\wedge d{\hat x}^\nu,\\[2mm]
G^{(L)}(x)&={1\over 2}
            \left(
            \ba{cc}
            0&0\\
            0&\lambda_aG^a_{\mu\nu}(x)d{\hat x}^\mu\wedge d{\hat x}^\nu\\
            \ea
            \right)\\[2mm]
G^{(R)}(x)&={1\over 2}
            \left(
            \ba{ccc}
            0&0&0\\
            0&\lambda_aG^a_{\mu\nu}(x)d{\hat x}^\mu\wedge d{\hat x}^\nu&0\\
            0&0&\lambda_aG^a_{\mu\nu}(x)d{\hat x}^\mu\wedge d{\hat x}^\nu\\
            \ea
            \right)
\label{eqn:a41}
\ea\ee
with the usual field strengths
\be\ba{cl}
{\vec f}_{\mu\nu}&=\pd_\mu {\vec A}_\nu-\pd_\nu{\vec A}_\mu
+g{\vec A}_\mu\times{\vec A}_\nu,\quad
f_{\mu\nu}^0=\pd_\mu B_\nu-\pd_\nu B_\mu,\\[2mm]
G_{\mu\nu}^a&=\pd_\mu G_\nu^a-\pd_\nu G_\mu^a
+g_sf_{abc}{G_\mu}^b{G_\nu}^c,
\label{eqn:a42}
\ea\ee
$f_{abc}$ being $SU(3)$ structure constants.
\par
The covariant derivative, $D_\mu H(x,y)$,
of the Higgs field (\ref{eqn:a38}) is given
by applying $D_\mu$ on $\phi$ and $\tphi$ in Eq.(\ref{eqn:a38})
with
\be\ba{cl}
D_\mu\phi&=
(\pd_\mu-i{g\over 2}{\vec\tau}\cdot{\vec A}_\mu-i{g'\over 2}
B_\mu)\phi\\[2mm]
D_\mu\tphi&=
(\pd_\mu-i{g\over 2}{\vec\tau}\cdot{\vec A}_\mu+i{g'\over 2}
B_\mu)\tphi.
\label{eqn:a43}
\ea\ee
\par
Now we determine the form of the matrix $C(y)$ from
the gauge invariance (\ref{eqn:a8}).
It turns out that
\be\ba{cl}
C(+)&=
\left(
\ba{cc}
c_L^{(l)}1_2&0\\
0&c_L^{(q)}1_2\otimes 1_3\\
\ea\right)\\[2mm]
C(-)&=
\left(
\ba{ccc}
c_R^{(e)}&0&0\\
0&c_R^{(d)}1_3&0\\
0&0&c_R^{(u)}1_3\\
\ea\right),
\label{eqn:a44}
\ea\ee
where $c_{L,R}^{(f)}, f=l,q,e,d,u$
are the constant matrices in the generation space.
The condition $C(y)M(y)=-M(y)C(-y)$ of Eq.(\ref{eqn:a9})
is satisfied by
$c_L^{(l)}a^{(e)}=-a^{(e)}c_R^{(e)},\hfill\break
c_L^{(q)}a^{(d)}=-a^{(d)}c_R^{(d)}$ and
$c_L^{(q)}a^{(u)}=-a^{(u)}c_R^{(u)}$.
The Higgs potential comes from the last term on the right-hand side of
Eq.(\ref{eqn:a27}).
\par
Finally we have to parametrize the
coupling-constants-matrix ${1\over 4g_y^2}\equiv {1\over 4G^2}
C_y^2$ in Eq.(\ref{eqn:a30}):
\be\ba{cl}
C_+^2&=
       \left(
       \ba{cc}
       1_2&0\\
       0&C_Q^21_2\otimes 1_3\\
       \ea\right)\\[2mm]
C_-^2&=
       \left(
       \ba{ccc}
       \delta^2&0&0\\
       0&\delta^2C_Q^21_3&0\\
       0&0&\delta^2C_Q^21_3\\
       \ea\right),
\label{eqn:a46}
\ea\ee
where we have introduced two more parameters $\delta^2$ and
$C_Q^2$.
The most general form of $C_-^2$ contains different parameters
$C_d^2$ and $C_u^2$ which are assumed to be equal to $C_Q^2$ in
Eq.(\ref{eqn:a46}).
By choosing
\be\ba{cl}
G^2&={g_s^2\over 2}N_g(1+\delta^2)C_Q^2,\\[2mm]
G^2&={g^2\over 4}N_g(1+3C_Q^2),\\[2mm]
G^2&={{g'}^2\over 8}N_g{\rm tr}[C_+^2Y_L^2+C_-^2Y_R^2]
={{g'}^2\over 4}N_g[(1+2\delta^2)+{1\over 3}C_Q^2(1+10\delta^2)],\\[2mm]
G^2&={\alpha^2\over 4}(1+\delta^2)[{\rm tr}(a^{(e)\dag}a^{(e)})
+3C_Q^2{\rm tr}(a^{(d)\dag}a^{(d)})+
3C_Q^2{\rm tr}(a^{(u)\dag}a^{(u)})],
\label{eqn:a47}
\ea\ee
we finally find that
\be\ba{cl}
\cL_B&=
-\sum_{y=\pm}{\rm tr}{1\over 4g_y^2}<\cF(x,y),\cF(x,y)>\\[2mm]
&=-{1\over 4}[G_{\mu\nu}^aG^{a,\mu\nu}
+{\vec f}_{\mu\nu}\cdot{\vec f}^{\mu\nu}
+f_{\mu\nu}^0f^{0,\mu\nu}]\\[2mm]
&\quad+(D_\mu\phi)^{\dag}(D^\mu\phi)-{\lambda\over 4}
(\phi^{\dag}\phi)^2+\mu^2(\phi^{\dag}\phi)+{\rm const.},
\label{eqn:a48}
\ea\ee
where we have
\be
\lambda=2\ep^2{{{\rm tr}(a^{(e)}a^{(e)\dag})^2
+3C_Q^2{\rm tr}(a^{(d)}a^{(d)\dag})^2
+3C_Q^2{\rm tr}(a^{(u)}a^{(u)\dag})^2}\over
{{\rm tr}(a^{(e)}a^{(e)\dag})
+3C_Q^2{\rm tr}(a^{(d)}a^{(d)\dag})
+3C_Q^2{\rm tr}(a^{(u)}a^{(u)\dag})}},
\label{eqn:a50}
\ee
with $\ep^2={\beta^2\over \alpha^2}$ \footnote{Sogami's $\lambda$
is four times ours and obtained by putting $\ep^2=1$
and $C_Q^2=1$.}
and, assuming $c_L^{(l)}=\varsigma, c_R^{(e)}=-\varsigma,
c_L^{(q)}=\varsigma, c_R^{(d)}=-\varsigma$ and
$c_R^{(d)}=-\varsigma$ for constant $\varsigma$,
\be
\mu^2=\ep^2\varsigma^2.
\label{eqn:a51}
\ee
\par
Equation (\ref{eqn:a48}) is nothing but the bosonic lagrangian in the
standard model with the following parametrization
for gauge coupling constants
\be
{g_s^2\over g^2}={{1+3C_Q^2}\over{2C_Q^2(1+\delta^2)}}
\label{eqn:a52}
\ee
and
\be
{{g'}^2\over g^2}=\tan^2{\theta_W}=
{1+3C_Q^2\over (1+2\delta^2)+{1\over 3}C_Q^2(1+10\delta^2)}
\label{eqn:a53}
\ee
where $\theta_W$ is Weinberg angle.
Sogami$^{10)}$ emphasized that
it is not necessary to imagine the two-sheeted world
as far as the derivation of the bosonic lagrangian (\ref{eqn:a48})
is concerned.
In fact, we shall see in the next section that
the same lagrangian (\ref{eqn:a48}) is obtained by the matrix method
without explicitly referring to the discrete space-time.
We, therefore, conclude that
our algebraic rule based on the algebra of functions over
the discrete space-time,
$X=M_4\times Z_2$,
defines only a convenient mathematical manipulation
consistent with the chiral nature of fermions
but the standard model itself is reconstructed solely on the continuous
manifold $M_4$.
Equations (\ref{eqn:a52}) and (\ref{eqn:a53})
were already derived in Ref.12) where additional parameter $c$
appeared due to an alternative choice of the chiral spinors,
leading to different equation for $\lambda$ than Eq.(\ref{eqn:a50}).
\par
If quarks and leptons contribute equally to the bosonic lagrangian,
$C_Q^2=1$,
and, moreover, $\delta^2$ is taken to be unity,
we obtain $SU(5)$ relations $g_s^2=g^2={5\over 3}{g'}^2$.
This is realized provided that $C_+^2$ and $C_-^2$
are unit matrices of dimensions 8 and 7, respectively,
which, therefore, should reflect $SU(5)$
symmetry in some sense.
As noted in Ref.12),
rough estimation for ${g_s^2\over g^2}\sim 4$ and
${{g'}^2\over g^2}\sim {1\over 3}$ at present energy
gives $C_Q^2\sim {1\over 14}$ and $\delta^2\sim {7\over 6}$.
If we retain only top quark contribution to the quartic coupling constant
(52) \footnote{
One can diagonalize all matrices $a^{(f)}, f=e,d,u,$
to estimate Eq.(\ref{eqn:a50})
provided Kobayashi-Maskawa matrix appears for the charged current
interactions.},
we would have $m_H\approx\ep\sqrt{2}m_t$,
where $m_H$ and $m_t$ denote
physical Higgs boson and top quark masses, respectively.
This is to be compared with
Sogami's prediction$^{10)}$
$m_H\approx\sqrt{2}m_t$ which is also
reported in
the famous paper by Connes and Lott in Ref.1)
where $\lambda$, given essentially
by replacing tr$(a^{(f)}a^{(f)\dag})^2$ in
Eq.(\ref{eqn:a50}) with
tr$(a^{(f)}a^{(f)\dag})^2-N_g^{-1}$(tr$a^{(f)}a^{(f)\dag})^2$,
vanishes for $N_g=1$.
The appearance of the parameter $\ep$
completely makes the mass relation ambiguous,
but
we expect that $\ep^2$ is of order unity.
\section*{\large \bf {{\mbox \S 5}
Matrix Method in the Chiral Space}}
\par
This section is essentially a repetition of
the previous section
using the matrix notation in the chiral space:
\be
\Psi(x)=
\left(
\ba{cl}
\psi_L(x)=\psi(x,+)\\
\psi_R(x)=\psi(x,-)\\
\ea
\right)
\label{eqn:a54}
\ee
and
\be
\cD_\mu(x)=
\left(
\ba{cc}
D_\mu^L(x)-{1\over 4}c_L\gamma_\mu&{i\over 4}\gamma_\mu H(x)\\
{i\over 4}\gamma_\mu H^{\dag}(x)&D_\mu^R(x)-{1\over 4}c_R\gamma_\mu\\
\ea
\right),
\label{eqn:a55}
\ee
where $D_\mu^L(x)=D_\mu(x,+)$, $D_\mu^R(x)=D_\mu(x,-)$,
$c_L=C(+)$, $c_R=C(-)$, $H(x)=H(x,+)$ and $H^{\dag}(x)=H(x,-)$.
\par
The assignment (\ref{eqn:a35})
then leads to, omitting the generation and color indices,
\be
\Psi=
\left(
\ba{cl}
l_L\\
q_L\\
e_R\\
d_R\\
u_R\\
\ea
\right),
\label{eqn:a56}
\ee
in terms of which the fermionic lagrangian (\ref{eqn:a34}) with
Eqs.(\ref{eqn:a34-1}) and (\ref{eqn:a34-2})
is rewritten as
\be
\cL_D=i{\bar\Psi}\gamma^\mu\cD_\mu\Psi,
\label{eqn:a57}
\ee
where $\cD_\mu$ is given by Eq.(\ref{eqn:a55})
with $D_\mu^{L,R}$ containing the unit matrix $1_{N_g}$.
We should insert the expressions for
$D_\mu^L(x)=\partial_\mu+\cA_\mu(x,+)$,
$D_\mu^R(x)=\partial_\mu+\cA_\mu(x,-),
c_L=C(+), c_R=C(-)$ and $H(x)=H(x,+)$
from Eqs.(\ref{eqn:a37}),
(\ref{eqn:a38}) and (\ref{eqn:a39}), respectively.
The condition $C(y)M(y)=-M(y)C(-y)$
is translated into the one
$c_LA+Ac_R=A^{\dag}c_L+c_RA^{\dag}$,
where $H(x,+)$ of Eq.(\ref{eqn:a38}) is rewritten as
the product
$H(x,+)=A\Phi_1(x)$
with
$\Phi_1(x)=\left(
         \ba{ccc}
         \phi(x)&0&0\\
         0&\phi(x)&0\\
         0&0&\tphi(x)\\
         \ea
         \right)$.
It eliminates the linear terms in $c_{L,R}$
from
the field strength$^{10)}$
\be
\cF_{\mu\nu}=[\cD_\mu,\cD_\nu].
\label{eqn:a58}
\ee
For later purpose we introduce the associated field strength$^{12)}$
\be
{\tilde \cF}_{\mu\nu}=
\sum_\alpha h_\alpha^2\Gamma_\alpha \cF_{\mu\nu}\Gamma^\alpha,
\label{eqn:a59}
\ee
where the sum over $\alpha$ runs over $S,V,A,T$ and $P$
corresponding to $\Gamma_\alpha=1,\gamma_\lambda,\gamma_5\gamma_\lambda,
\hfill\break
\sigma_{\lambda\rho}={i\over 2}[\gamma_\sigma,\gamma_\rho]$
and $i\gamma_5$,
respectively.
Putting
\be\ba{cl}
\sum_\alpha h_\alpha^2\Gamma_\alpha\gamma_\mu\Gamma^\alpha&=
(h_S^2-2h_V^2-2h_A^2+h_P^2)\gamma_\mu\equiv -{2\over 3}
\alpha^2\gamma_\mu,\\[2mm]
\sum_\alpha h_\alpha^2\Gamma_\alpha\sigma_{\mu\nu}\Gamma^\alpha&=
(h_S^2-4h_T^2-h_P^2)\sigma_{\mu\nu}\equiv
{2\over 3}\beta^2\sigma_{\mu\nu},\\[2mm]
\sum_\alpha h_\alpha^2\Gamma_\alpha \Gamma^\alpha&=
h_S^2+4h_V^2-4h_A^2+12h_T^2-h_P^2\equiv 1,
\label{eqn:a60}
\ea\ee
we define the bosonic lagrangian
\be
\cL_B=-{1\over 32G^2}{\rm Tr}[C^2\gamma_0\cF_{\mu\nu}^{\dag}
\gamma_0{\tilde\cF}^{\mu\nu}],
\label{eqn:a61}
\ee
where Tr means the trace over Dirac matrices,
the 2$\times $2 matrices in the chiral space and the internal
symmetries
and
$C^2=
\left(
\ba{cc}
C_+^2&0\\
0&C_-^2\\
\ea
\right)$.
It is straightforward to show that
Eq.(\ref{eqn:a61})
yields precisely the same lagrangian (\ref{eqn:a48}) for the same relations
(\ref{eqn:a47})
with the parametrizations (\ref{eqn:a50}),
(\ref{eqn:a51}), (\ref{eqn:a52}) (for
$c_L=\varsigma 1_8$ and $c_R=-\varsigma 1_7$)
and (\ref{eqn:a53}).
In other words, the introduction of the associated field strength
(\ref{eqn:a59})
with Eq.(\ref{eqn:a60})
reflects the independence of the inner products of the
one-form $\chi$ and two-form $\chi\wedge\chi$
as exemplified in Eqs.(\ref{eqn:a23})
and (the last equation of) (\ref{eqn:a31}) with the same meaning
of the parameters $\alpha^2$ and $\beta^2$.
\section*{\large \bf {{\mbox \S 6} Summary}}
\par
To conclude we have been able to reconstruct the standard model
within the framework of the modified formalism
of the non-commutative differential geometry
and to reinterpret the modified formalism
in relation to Sogami's method$^{10)}$.
An obvious
next question is how to extend the present formalism
so as to describe more than one Higgs fields.
\par
It is believed among NCG-minded people that
NCG gives a constrained standard model$^{15)}$.
This conclusion depends on the choice of the starting,
involutive algebra and Connes' definition of
\YMH lagrangian through the Dixmier trace.
In contrast,
we reproduced the standard model without any constraints among
the tree-level parameters.
Our formalism parallels the ordinary differential geometry
as closely as possible.
Nonetheless, our reconstruction strongly depends on
the pattern of existence of fermions.
\par
The biggest departure from the
ordinary differential geometry is the introduction
of the extra one-form basis $\chi$
which does not vanish upon taking the wedge product
and allows one to consider an algebraic sum of
matrices of different types, namely, $A+B\chi$ (see below Eq.(\ref{eqn:a11}).)
The latter aspect is only a
convenient mathematical magic to treat gauge and Higgs fields
in a unified way as a single, generalized one-form (\ref{eqn:a14})
where the shifted Higgs field appears.
If we make use of the 2$\times$2 matrix rep$^{14)}$
considering the direct sum of Hilbert spaces ${\cal H}_+\oplus{\cal H}_-$,
$\chi$ twisted-commutes with bosonic and fermionic
fields as shown in the Appendix A\footnote{Based on $Z_2$-graded algebra of
Ref.14) we can assume that $\chi$ simply commutes with
bosonic and fermionic variables.}.
In this case,
we may write Eq.(\ref{eqn:a16})
as
\be
\cD(x)\psi(x)=(D(x)+iC\chi+H(x)\chi)\psi(x),
\label{eqn:a62}
\ee
where\footnote{Cartan's structure equation of the connection
one-form $\cA(x)$
determines the curvature 2-form in this notation
$$
\cF(x)=d\cA(x)+\cA(x)\wedge\cA(x),
$$
where we put $\cD(x)=d+\cA(x)$ and adopt the convention$^{14)}$
of regarding $f=(f_1,f_2), f=D,C,H$, as elements of $Z_2$-graded algebra
so that $f\chi=\chi f$.
This is the matrix form of Eq.(\ref{eqn:a25}).}
\be\ba{cl}
\psi(x)&=\left(
        \ba{cl}
        \psi(x,+)\\
        \psi(x,-)\\
        \ea
        \right),\;\;
D(x)=\left(
     \ba{cc}
     D(x,+)&0\\
     0&D(x,-)\\
     \ea
     \right),\\[2mm]
C&=\left(
  \ba{cc}
  C(+)&0\\
  0&C(-)\\
  \ea
  \right),\;\;
H(x)=\left(
     \ba{cc}
     0&H(x,+)\\
     H(x,-)&0\\
     \ea
     \right).
\label{eqn:a64}
\ea\ee
In this respect
we recall that
Sogami's generalized covariant derivative
(\ref{eqn:a55}) is rewritten as$^{13)}$
\be
\cD(x)\equiv\cD_\mu(x) d{\hat x}^\mu=D(x)+iC\chi+\tH(x)\chi,
\label{eqn:a65}
\ee
where
\be
\chi={i\over 4}\gamma_\mu d{\hat x}^\mu
\label{eqn:a66}
\ee
acts on the spinor from the left and
\be
D=\left(
  \ba{cc}
  D_\mu^L d{\hat x}^\mu&0\\
  0&D_\mu^R d{\hat x}^\mu
  \ea
  \right),\;
C=\left(
  \ba{cc}
  c_L&0\\
  0&c_R\\
  \ea
  \right),\;
\tH=\left(
    \ba{cc}
    0&H\\
    H^{\dag}&0\\
    \ea
    \right).
\label{eqn:a67}
\ee
Therefore, Sogami's method$^{10)}$
provides us with a concrete realization of
the mysterious symbol $\chi$.
It is important to remember, however, that
our formalism does not presuppose a concrete
realization of $\chi$. Hence, even fermionic variable
is multiplied by $\chi$ from both sides.
\par
Last but not least we quote Ref.16) which precedes various works$^{2)-9)}$
on NCG approach to particle models.
The authors in Ref.16) introduced the matrix derivation and
proposed particle models which contain Higgs bosons
belonging to the adjoint rep.
Our allowance of taking an algebraic sum of
matrices of different types,
or equivalently,
our introduction of the concept of $Z_2$-grading in the
2$\times$2 matrix rep\footnote{The concept of $Z_2$-grading
in the matrix formulation of NCG was first introduced in Ref.3)
where the symbol $\chi$ was not yet introduced.}
fits to the fact that
Higgs field in the theory
belongs to any unitary rep as far as it couples to fermions.
\par
\vspace{2.5cm}
\centerline{\sf Acknowledgements}
\vspace{5mm}
{\sf The authors are grateful
to Professor J. Iizuka, Professor H. Kase and
Professor M. Tanaka for
informing some references, useful discussions and continuous encouragement.}
\vfill
\newpage
\begin{center}
{\large\bf Appendix A}
\end{center}
\par
We shall add some comments on our interpretation of Eqs.(\ref{eqn:a5})
and (\ref{eqn:a11}).
\par
As we showed in Ref.14), the relation
(\ref{eqn:a11})
should be regarded as a calculational rule but not as a matrix equation.
To see this let us employ the
2$\times$2 matrix rep$^{14)}$
\bea
f(x)&=&\left(
    \ba{cc}
    f_1(x)=f(x,+)&0\\
    0&f_2(x)=f(x,-)\\
    \ea
    \right),\;\;{\rm for}\;{\rm even}\;{\rm function}\;f(x,y),
    \nn
    \\[2mm]
g(x)&=&\left(
    \ba{cc}
    0&g_1(x)=g(x,+)\\
    g_2(x)=g(x,-)&0\\
    \ea
    \right),\;\;{\rm for}\;{\rm odd}\;{\rm function}\;g(x,y),
    \nn\\[2mm]
M&=&\left(
    \ba{cc}
    0&M_1=M(+)\\
    M_2=M(-)&0\\
    \ea
    \right),\;\;{\rm for}\;{\rm odd}\;{\rm function}\;M(y).
    \nn
\eea
Then Eq.(\ref{eqn:a6}) is brought to
$$
d_\chi f(x)=[Mf(x)-(-1)^{\partial f}f(x)M]\chi,
\eqno(A\cdot 1)
$$
if we write
\begin{eqnarray*}
d_\chi f(x)&=&
d_\chi\left(
      \ba{cc}
      f_1(x)&0\\
      0&f_2(x)\\
      \ea
      \right)=\left(
              \ba{cc}
              0&d_\chi f_1(x)\\
              d_\chi f_2(x)&0\\
              \ea
              \right)
                 \\[2mm]
             &=&\left(
               \ba{cc}
               0&\partial_\chi f_1(x)\\
               \partial_\chi f_2(x)&0\\
               \ea
               \right)\chi=(\partial_\chi f(x))\chi,
                 \\[2mm]
d_\chi g(x)&=&
d_\chi\left(
      \ba{cc}
      0&g_1(x)\\
      g_2(x)&0\\
      \ea
      \right)=\left(
              \ba{cc}
              d_\chi g_1(x)&0\\
              0&d_\chi g_2(x)\\
              \ea
              \right)\\[2mm]
              &=&
              \left(
              \ba{cc}
              \partial_\chi g_1(x)&0\\
              0&\partial_\chi g_2(x)\\
              \ea
              \right)\chi=(\partial_\chi g(x))\chi.
\end{eqnarray*}
Then Eq.(A$\cdot$ 1) makes sense as matrix equation yielding
\begin{eqnarray*}
d_\chi f_1(x)&=&[M_1f_2(x)-f_1(x)M_1]\chi,\;\;
d_\chi f_2(x)=[M_2f_1(x)-f_2(x)M_2]\chi,\\[2mm]
d_\chi g_1(x)&=&[M_1g_2(x)+g_1(x)M_2]\chi,\;\;
d_\chi g_2(x)=[M_2g_1(x)+g_2(x)M_1]\chi.
\end{eqnarray*}
The reason for writing $d_\chi f$ and $d_\chi g$
in the above way lies in the fact that $d_\chi$ changes the grade.
On the other hand,
the Leibniz rule (\ref{eqn:a10}) written for $\partial_\chi$
is combined into
$$
\partial_\chi(f(x)g(x))=(\partial_\chi f(x))g(x)
+(-1)^{\partial f}f(x)\partial_\chi g(x).
\eqno(A\cdot 2)
$$
The Leibniz rule (\ref{eqn:a10})
then reads
$$
d_\chi(f(x)g(x))=(d_\chi f(x))\tau_1g(x)\tau_1
+(-1)^{\partial f}f(x)d_\chi g(x).
\eqno(A\cdot 3)
$$
Equation(A$\cdot$2)
implies Eq.(A$\cdot$3)
if
$$
f(x)\chi=\chi \tau_1f(x)\tau_1,\;\;f(x):\;{\rm
even}\;{\rm or}\;{\rm odd},
\eqno(A\cdot 4)
$$
which is a matrix equation\footnote{A matrix equation
involving $\chi$ should always take the form $A\chi=B\chi$,
implying the usual matrix equation $A=B$.
{}From Eq.(A$\cdot$4) $\chi$ can not be a ``scalar'' in the matrix
multiplication law. If, on the other hand, we regard $f=(f_1,f_2)$ as elements
of
a $Z_2$-graded algebra, we simply have$^{14)}$ $f\chi=\chi f$ which was
employed in defining the curvature 2-form in the footnote on p.20.}.
Here the matrix $\tau_1$ plays a role of exchanging the indices
1$\leftrightarrow $2:
\begin{eqnarray*}
\tau_1\left(
      \ba{cc}
      f_1&0\\
      0&f_2\\
      \ea
      \right)\tau_1&=&\left(
                     \ba{cc}
                     f_2&0\\
                     0&f_1\\
                     \ea
                     \right),\\[2mm]
\tau_1\left(
      \ba{cc}
      0&g_1\\
      g_2&0\\
      \ea
      \right)\tau_1&=&\left(
                     \ba{cc}
                     0&g_2\\
                     g_1&0\\
                     \ea
                     \right).
\end{eqnarray*}
We can recover
(A$\cdot$4) if the following relations
are assumed to be valid:
$$
f_1(x)\chi=\chi f_2(x),\;\;f_2(x)\chi=\chi f_1(x),\;\;
f:\;{\rm even}\;{\rm or}\;{\rm odd},
\eqno(A\cdot 5)
$$
which are nothing but Eq.(\ref{eqn:a10}).
Equations (A$\cdot$5)
are not matrix equations.
\par
The same is true also for
Eq.(\ref{eqn:a5}).
The reasoning is the same as above.
Putting
\begin{eqnarray*}
\psi(x)&=&\left(
        \ba{cl}
        \psi_1(x)=\psi(x,+)\\
        \psi_2(x)=\psi(x,-)\\
        \ea
        \right)
        \\[2mm]
C&=&\left(
    \ba{cc}
    C_1=C(+)&0\\
    0&C_2=C(-)\\
    \ea
    \right)
\end{eqnarray*}
we rewrite the third equation of Eq.(\ref{eqn:a2}) as
$$
d_\chi\psi(x)=[M\psi(x)+iC\psi(x)]\chi,
$$
where we should write
$$
d_\chi\psi(x)=\left(
              \ba{cl}
              d_\chi\psi_1(x)\\
              d_\chi\psi_2(x)\\
              \ea
              \right)
              =\left(
               \ba{cl}
               \partial_\chi\psi_1(x)\\
               \partial_\chi\psi_2(x)\\
               \ea
               \right)\chi=(\partial_\chi\psi(x))\chi.
$$
Similarly,
Eq.(\ref{eqn:a3}) written for the operator
$\partial_\chi$
reads
$$
\partial_\chi(f(x)\psi(x))
=(\partial_\chi f(x))\psi(x)+(-1)^{\partial f}f(x)\partial_\chi\psi(x).
\eqno(A\cdot 6)
$$
The Leibniz rule for the operator $d_\chi$ is obtained as
$$
d_\chi(f(x)\psi(x))
=(d_\chi f(x))\tau_1\psi(x)+(-1)^{\partial f}f(x)d_\chi\psi(x).
\eqno(A\cdot 7)
$$
Hence, we have
$$
\psi(x)\chi=\chi\tau_1\psi(x),\;\;\tau_1\left(
                                        \ba{cl}
                                        \psi_1\\
                                        \psi_2\\
                                        \ea
                                        \right)
                                        =\left(
                                         \ba{cl}
                                         \psi_2\\
                                         \psi_1\\
                                         \ea
                                         \right).
\eqno(A\cdot 8)
$$
This matrix equation is also
interpreted as indicating a mere mathematical rule
$$
\chi\psi_1(x)=\psi_2(x)\chi,\;\;\chi\psi_2(x)=\psi_1(x)\chi,
\eqno(A\cdot 9)
$$
which is nothing but Eq.(\ref{eqn:a5}).
This allows us to consistently
apply the matrix multiplication rule
in our formalism.
\par
\par
\begin{center}
{\large\bf Appendix B}
\end{center}
\par
To compare with the formalism of Ref.11),
we insert here an additional remark concerning the gauge status of Higgs
field.
To be more precise we shall rederive Eq.(\ref{eqn:a15-1})
in a different way more close to that of Ref.11).
\par
Consider $\psi(x)$ in the Appendix A as a section of spinor bundle
$S$.
Locally it is expanded in terms of the basis $\{{\bf E}_K(x)\}$
of fibres $S_x=\{S_x^+, S_x^-\}$, to be called local frame fields:
$$
\psi(x)=\sum_{K}{\bf E}_K(x)\psi^K(x)
\equiv\left(
\ba{cl}
\sum_{k=1}^{k=m}{\bf e}_k(x,+)\psi_1^k(x)\\
\sum_{l=1}^{l=n}{\bf e}_l(x,-)\psi_2^l(x)\\
\ea
\right).
\eqno(B\cdot 1)
$$
The right-hand side may be calculated through the matrix multiplication rule
by
representing
$$
{\bf E}_K(x)=\left(
             \ba{cc}
             {\bf e}_k(x,+)&0\\
             0&{\bf e}_l(x,-)\\
             \ea
             \right)
$$
and performing the sum over the indices $\{k,l\}$ after multiplication
with
$$
\psi^K(x)=\left(
           \ba{cl}
           \psi_1^k(x)\\
           \psi_2^l(x)\\
           \ea
           \right).
$$
In what follows we follow this convention.
The fibre spaces
$S_x^+$ and $S_x^-$ may have different dimensions, $m$ and $n$,
respectively,
with corresponding basis
$\{{\bf e}_k(x,+)\}_{k=1,\cdots,m}$
and
$\{{\bf e}_l(x,-)\}_{l=1,\cdots,n}$.
The index $K$ takes on the values 1,2,$\cdots,(m+n)$.
The covariant derivative in the discrete direction,
denoted $\nabla_\chi$ here,
is defined by
$$
\nabla_\chi\psi(x)=\sum_{K}[\nabla_\chi {\bf E}_K(x).\tau_1\psi^K(x)
+{\bf E}_K(x)d_\chi\psi^K(x)],
\eqno(B\cdot 2)
$$
where we have used the Leibniz rule (A$\cdot$7).
Since $\nabla_\chi {\bf E}_K(x)$ can be written as a linear combination of
the basis ${\bf E}_K(x)$,
we put
$$
\nabla_\chi {\bf E}_K(x)=\sum_{L}{\bf E}_L(x)\Phi^L_{\;K}(x)\chi
\eqno(B\cdot 3)
$$
where the connection form $\Phi(x)=(\Phi^L_{\;K}(x))$
comprises an odd function:
$$
\Phi(x)=(\Phi^L_{\;K}(x))=\left(
                      \ba{cc}
                      0&\Phi_1(x)\\
                      \Phi_2(x)&0\\
                      \ea
                      \right).
$$
Thus we have
\begin{eqnarray*}
&&
\sum_K\nabla_\chi {\bf E}_K(x)\tau_1\psi^K(x)\\[2mm]
&&\!\!\!\!\!\!\!\!\!=
 \left(
 \ba{cl}
 \sum_{l=1}^{l=m}{\bf e}_l(x,+)\sum_{k=1}^{k=n}
 {\Phi_1^l}_{\;k}(x)\psi_2^k(x)\\
 \sum_{l=1}^{l=n}{\bf e}_l(x,-)\sum_{k=1}^{k=m}
 {\Phi_2^l}_{\;k}(x)\psi_1^k(x)\\
 \ea
 \right)\chi.
\end{eqnarray*}
The extra exterior derivative $d_\chi\psi^K(x)$ is given by
$$
d_\chi \psi^K(x)=\sum_{L}(M^K_{\;\;L}+iC^K_{\;L})\psi^L(x)\chi
$$
where the matrices $M=(M^K_{\;\;L})$ and
$C=(C^K_{\;\;L})$ are
the same ones as defined in the previous Appendix A
so that
\begin{eqnarray*}
&&\sum_{K}{\bf E}_K(x)d_\chi\psi^K(x)=\\[2mm]
&&\!\!\!\!\!\!\!\!\!\!\!\!
 \left(
 \ba{cl}
 \sum_{l=1}^{l=m}{\bf e}_l(x,+)
 \sum_{k=1}^{k=m}i{C_1^l}_{\;k}(x)\psi_1^k(x)+\sum_{l=1}^{l=m}{\bf e}_l(x,+)
 \sum_{k=1}^{k=n}{M_1^l}_{\;k}(x)\psi_2^k(x)\\
 \sum_{l=1}^{l=n}{\bf e}_l(x,-)\sum_{k=1}^{k=m}{M_2^l}_{\;k}(x)\psi_1^k(x)+
 \sum_{l=1}^{l=n}{\bf e}_l(x,-)\sum_{k=1}^{k=n}i{C_2^l}_{\;k}(x)\psi_2^k(x)\\
 \ea
 \right)\chi.
\end{eqnarray*}
Substituting them back into Eq.(B$\cdot 2$) we find that
\begin{eqnarray*}
\nabla_\chi\psi(x)&=&\sum_{L,K}{\bf E}_L(x)H^L_{\;K}(x)\psi^K(x)\chi
                   =(H(x)+iC)\psi(x)\chi\\[2mm]
&&\!\!\!\!\!\!\!\!\!\!\!\equiv
\left(
\ba{cl}
\sum_{k=1}^{k=m}{\bf e}_k(x,+)(i\sum_{l=1}^{l=m}{C_1^k}_{\;l}(x)\psi_1^l(x)+
\sum_{l=1}^{l=n}{H_1^k}_{\;l}(x)\psi_2^l(x))\\
\sum_{l=1}^{l=n}{\bf e}_l(x,-)(i\sum_{l=1}^{l=n}{C_2^k}_{\;l}(x)\psi_2^l(x)+
\sum_{k=1}^{k=m}{H_2^l}_{\;k}(x)\psi_1^k(x))\\
\ea
\right)\chi,
\end{eqnarray*}
where we have
put
$H=\Phi+M$.
\par
We finally determine the transformation property of
the connection form
$\Phi(x)$ under the transformation of the local frame fields:
\begin{eqnarray*}
{^g{\bf E}}_K(x)&=&\sum_{L}{\bf E}_L(x){(g^{-1})}^L_{\;K}(x)\\[2mm]
&=&
  \left(
  \ba{cc}
  \sum_{l=1}^{l=m}{\bf e}_l(x,+){({g_1^{-l}})}^l_{\;k}(x)&0\\
  0&\sum_{l=1}^{l=n}{\bf e}_l(x,-){({g_2^{-l}})}^l_{\;k}(x)\\
  \ea
  \right),
\end{eqnarray*}
where
$$
g(x)=(g^L_{\;K}(x))=
     \left(
     \ba{cc}
     g_1(x)&0\\
     0&g_2(x)\\
     \ea
     \right)
$$
is the gauge transformation function\footnote{Strictly speaking,
we should write $\rho(g(x))$ in place of $g(x)$.
We can neglect the difference as far as Higgs field is concerned.}.
Noting Eq.(A$\cdot$3) it follows that
\begin{eqnarray*}
\nabla_\chi\, ^g{\bf E}_K(x)&=&
\sum_{L}[\nabla_\chi{\bf E}_L(x).\tau_1{({g^{-1}})}^L_{\;K}(x)\tau_1
+{\bf E}_L(x).d_\chi{({g^{-1}})}^L_{\;K}(x)]\\[2mm]
&=&\sum_{L,P}{\bf E}_L(x)\Phi^L_{\;P}(x)\chi\tau_1{(g^{-1})}^P_{\;K}(x)\tau_1
+\sum_{L}{\bf E}_L(x).d_\chi{(g^{-1})}^L_{\;K}(x).
\end{eqnarray*}
Defining the transformed connection form by
$$
\nabla_\chi\, ^g{\bf E}_K(x)=\sum_{L}\, ^g{\bf E}_L(x)\,
{^g\Phi}^L_{\;K}(x)\chi
=\sum_{L,P}{\bf E}_L(x){({g^{-1}})}^L_{\;P}(x)\,{^g\Phi}^P_{\;K}(x)\chi
$$
and noting
\begin{eqnarray*}
&&\sum_{L,P}{\bf E}_L(x)\Phi^L_{\;P}(x)\chi\tau_1
{(g^{-1})}^P_{\;K}(x)\tau_1
=\sum_{L,P}{\bf E}_L(x)\Phi^L_{\;P}(x){(g^{-1})}^P_{\;K}(x)\chi\\[2mm]
&&\!\!\!\!\!\!\!\!\!\!\!\!\!\!\!\!\!\!\!\!\!\!\!=\left(
 \ba{cc}
 0&\!\!\!\!\!\!\!\!\!\!\sum_{l=1}^{l=m}{\bf e}_l(x,+)\sum_{p=1}^{p=n}
 {\Phi_1^l}_p(x){({g_2}^{-1})}^p_{\;k}(x)\\
 \sum_{l=1}^{l=n}{\bf e}_l(x,-)\sum_{p=1}^{p=m}
 {\Phi_2^l}_p(x){({g_1}^{-1})}^p_{\;k}(x)&\!\!\!\!\!\!\!\!\!\!0\\
 \ea
 \right)\chi\\[3mm]
&&\sum_{L,P}{\bf E}_L(x){({g^{-1}})}^L_{\;P}\,{^g\Phi}^P_{\;K}(x)\\[2mm]
&&\!\!\!\!\!\!\!\!\!\!\!\!\!\!\!\!\!\!\!\!\!\!\!=
 \left(
 \ba{cc}
 0&\!\!\!\!\!\!\!\!\!\!\sum_{l=1}^{l=m}{\bf e}_l(x,+)\sum_{p=1}^{p=m}
 {(g_1^{-1})}^l_{\;p}(x)\,{^g\Phi_1}^p_{\;k}(x)\\
 \sum_{l=1}^{l=n}{\bf e}_l(x,-)\sum_{p=1}^{p=n}
 {(g_2^{-1})}^l_{\;p}(x)\,{^g\Phi_2}^p_{\;k}(x)&\!\!\!\!\!\!\!\!\!\!0\\
 \ea
 \right)\\[3mm]
&&\sum_{L}{\bf E}_L(x){(d_\chi g^{-1})}^L_{\;K}\\[2mm]
&&\!\!\!\!\!\!\!\!\!\!\!\!\!\!\!\!\!\!\!\!\!\!\!=
 \left(
 \ba{cc}
 0&\sum_{l=1}^{l=m}{\bf e}_l(x,+)
 {(d_\chi g_1^{-1})}^l_{\;k}(x)\\
 \sum_{l=1}^{l=n}{\bf e}_l(x,-)
 {(d_\chi g_2^{-1})}^l_{\;k}(x)&0\\
 \ea
 \right),
\end{eqnarray*}
we get
\begin{eqnarray*}
^g\Phi_1(x)&=&g_1(x)\Phi_1(x)g_2^{-1}+g_1(x)\partial_\chi g_2(x),\\[2mm]
^g\Phi_2(x)&=&g_2(x)\Phi_1(x)g_1^{-1}+g_2(x)\partial_\chi g_1(x),\\[2mm]
\end{eqnarray*}
where we have used Eq.(A$\cdot 4$) and the orthonormality:
\begin{eqnarray*}
{\bf e}_k(x,+)\cdot{\bf e}_l(x,+)&=&\delta_{kl},\;\;
k,l=1,2,\cdots,m,\\[2mm]
{\bf e}_k(x,-)\cdot{\bf e}_l(x,-)&=&\delta_{kl},\;\;
k,l=1,2,\cdots,n.
\end{eqnarray*}
In the 2$\times$2 matrix rep the result is converted into
$$
^g\Phi(x)=g(x)\Phi(x)g^{-1}(x)+g(x)\partial_\chi g^{-1}(x).
$$
This is identical with Eq.(\ref{eqn:a15-1}) in the text.
Note that, in general,
$m\not=n$ in our formalism and
the connection form $\Phi(x)$ of
Eq.(B$\cdot 3$), or its {\it back-shifted} Higgs
field $H(x)$
can never been related to the metric
${\bf E}_K(x)\cdot{\bf E}_L(x)=\delta_{KL}$.
Consequently,
the conclusion of Ref.11) concerning the unitarity
of Higgs field is not applicable to the present formalism.
\vfill
\newpage
\begin{center}
       {\large\bf References}
\end{center}
\par
 \begin{tabbing}
xxxx\=xxxxxxxxxxxxxxxxxxxxxxxxxxxxxxxxxxxxxxxxxxxxxxxxxxxxxxxxxxxxxxx\=\kill
\ 1)\> A.~Connes, $\quad$in {\it The Interface of Mathematics
       and Particle Physics}, edited by\\
    \> $\qquad\qquad\qquad$ D.~Quillen, G.~Segal and S.~Tsou (Clarendon
       Press, Oxford, 1990),\\
    \> $\qquad\qquad\qquad$ pp.9-48.\\
    \> A.~Connes and J.~Lott, Nucl. Phys. {\bf B}(Proc. Suppl.)
       {\bf 18}(1990), 29.\\
\ 2)\> D.~Kastler, Reviews in Math. Phys. {\bf 5}(1993), 477.\\
\ 3)\> R.~Coquereaux, G.~Esposito-Farese and G.~Vaillant, Nucl. Phys.
       {\bf B353}(1991), 689.\\
    \> R.~Coquereaux, R.~H{\"a}{\ss}ling, N.~A.~Papadopoulos,
       and F.~Scheck,\\
    \> \MP {\bf A7} (1992), 2809.\\
    \> See also, S. Naka and E. Umezawa, \PTP {\bf 92}(1994), 189.\\
\ 4)\> A.~H.~Chamseddine, G.~Felder and J.~Fr\"olich, Nucl. Phys.
       {\bf B395}(1993), 672.\\
\ 5)\> N.~A.~Papadopoulos, J. Plass and F.~Scheck, \PL {\bf 324}(1994), 380.\\
\ 6)\> J. C. Varilly and J. M. Gracia-Bondia,
       Jour. Geom. Phys. {\bf 12}(1995), 223.\\
\ 7)\> T. Sch\"ucker and J.-M. Zylinski,
       Jour. Geom. Phys. {\bf 16}(1995), 207.\\
\ 8)\> A.~Sitarz, Phys. Lett. {\bf B308}(1993), 311;$\;$
       Jour. Geom. Phys. {\bf 15}(1995), 123.\\
    \> Sitarz' formalism was applied to the standard model by\\
    \> H.-G. Ding, H.-Y. Guo, J.-M. Li and K. Wu, Zeit. f. Phys.
       {\bf C64}(1994), 521.\\
\ 9)\> K.~Morita and Y.~Okumura, Prog. Theor. Phys.{\bf 91}(1994), 959.\\
\ 10)\> I.~S.~Sogami, Prog. Theor. Phys. {\bf 94}, (1995), 117.\\
\ 11)\> G. Konisi and T. Saito, Prog. Theor. Phys.{\bf 93}(1995), 1093.\\
\ 12)\> K.~Morita, Prog. Theor. Phys. {\bf 94}, (1995), 125.\\
\ 13)\> K.~Morita, Y.~Okumura and M.~Tanaka-Yamawaki,\\
     \> $\qquad\qquad\qquad$
        to appear in Prog. Theor. Phys. {\bf 94}, No.3(1995).\\
\ 14)\> J.~Iizuka, H.~Kase, K.~Morita, Y.~Okumura and M.~Tanaka-Yamawaki,\\
       \> $\qquad\qquad\qquad$Prog. Theor. Phys.{\bf 92}(1994), 397.\\
\ 15)\> See, for instance,
        A.~H.~Chamseddine and J.~Fr\"olich, \PL {\bf B314}(1993), 308.\\
\ 16)\> R.~D-Violette, R.~Kerner and J.~Madore, \jmp {\bf 31}(1990), 316,323
        and \\
     \> references therein.\\
 \end{tabbing}
\end{document}